\documentclass[11pt]{article}
\usepackage{graphicx}
\usepackage{epstopdf}
\usepackage{amssymb,bm}
\usepackage{bbm}
\usepackage{amsmath}
\usepackage{amsfonts}
\usepackage{hhline}
\usepackage[margin=1.0in]{geometry}
\usepackage{cite}
\usepackage[shortlabels]{enumitem}
\usepackage{blindtext}
\usepackage{setspace}
\usepackage{color}
\usepackage[font=small,labelfont=bf]{caption}
\usepackage{floatrow}
\usepackage{wrapfig}
\newfloatcommand{capbtabbox}{table}[][\FBwidth]

\allowdisplaybreaks

\usepackage{amsthm}

\usepackage{chngcntr}
\usepackage{hyperref}

\counterwithout{theorem}{section}
\counterwithout{definition}{section}
\counterwithout{lemma}{section}
\counterwithout{remark}{section}
\counterwithout{assumption}{section}
\counterwithout{proposition}{section}
\counterwithout{corollary}{section}
\counterwithout{claim}{section}
\AtBeginDocument{
  \label{CorrectFirstPageLabel}
  
}

\begin{document}
\title{Success of machine learning algorithms in dynamical mass measurements of galaxy clusters}
\author{Muhammad Haider Abbas}
\date{\today}
\maketitle
\begin{abstract}
In recent years, machine learning (ML) algorithms have been successfully employed in Astronomy for analyzing and interpreting the data collected from various surveys. The need for new robust and efficient data analysis tools in Astronomy is imminently growing as we enter the new decade. Astronomical data sets are growing both in size and complexity at an exponential rate and ML methodologies can revolutionize our ability to interpret observations and provide new means of discovery. In this essay we focus on recent success of ML algorithms in predicting the dynamical mass of galaxy clusters. We discuss the results of the study performed by Ho et al. \cite{Ho} and their implications, where it was found that ML algorithms outperform conventional statistical methods and can offer a robust and accurate tool for dynamical mass estimation.
\end{abstract}
\section{Introduction}
\begin{wrapfigure}{R}{0.35\textwidth}
    \includegraphics[width=0.9\textwidth]{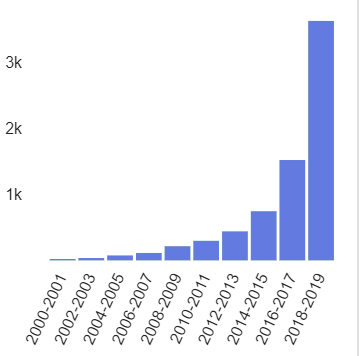}
  \caption{\footnotesize ADS papers that include ``machine learning'' in their abstracts. The figure shows an exponential growth with more than 3000 peer reviewed articles between 2018 and 2019 \cite{ADS}.}
  \label{fig0}
\end{wrapfigure}

The fields of Astronomy and observational Cosmology are entering an era of unprecedented data production and traffic as data sets continuously grow both in size and complexity. Although the ever-growing size in astronomical data sets faces challenges not directly addressable by advances in data science, e.g. with regard to data storage and transfer, it is more often than not the complexity of said data sets that makes them a rich reservoir for the application of ML algorithms. In recent years astronomers have used various ML tools to harvest novel information from astronomical data sets, and have done so with an unmatched level of efficiency in comparison to conventional paradigms. Figure \ref{fig0} shows the number of Astrophysics Data System (ADS) \cite{ADS} peer reviewed papers containing ``machine learning'' in their abstracts between the years 2000 and 2019 with more than 3000 papers between 2018 and 2019. Both supervised and unsupervised ML algorithms are used in Astronomy depending on the task at hand, and arguably unsupervised algorithms are more crucial in science as they can yield new discoveries unknown to us previously. To name a few, supervised ML algorithms such as Support Vector Machine (SVM) and Random Forest (RF) are widely used in Astronomy in classification and regression tasks, see for example \cite{svm}. Unsupervised algorithms such as Hierarchical Clustering and K-means are used to identify different clusters in the data set, see for example \cite{ma} and \cite{kmean}. Furthermore, Convolutional Neural Networks (CNNs) are very popular in studying strong and weak gravitational lensing maps, for example \cite{cnn1,cnn2}.

Developing a robust data analysis tool kit for next generation Astronomy appropriate to the exponentially growing data sets is an ongoing interdisciplinary endeavor. Although ML techniques inarguably accelerated discovery in Astronomy, criticism is unavoidable and in this case, it is usually the lack on interpretability of ML algorithms which makes part of the community sceptical. This is more so valid in Physics (rather than Computer Science) where the underlying mechanism or pattern an algorithm uses might itself be related to physical laws; hence to understand how a pattern emerges is essential for understanding the physics\footnote{The applicability or rather the need for ML algorithms in science is dependant on whether scientists can make use of a more sophisticated tool to analyze the date. This is pronounced in Astronomy where data sets pose challenges only addressable by means beyond conventional statistical methods, for further elaboration on why this is the case in Astronomy, see \cite{review}.}. However, to peek inside the box is not impossible and is an active area of research. This can be achieved by e.g. studying which input parameters are most crucial for the output \cite{simo}, or those that maximise activation in layers of a neural network \cite{ntamp}. Regardless of how deep we can delve into the know-how of a deep neural network, the current state of the art is very promising and in the next section, we elaborate on the need for employing ML algorithms in Astronomy.
\section{The need for ML in the era of big astronomical data}
An example of the surveys that introduced astronomy to the era of big data is the Sloan Digital Sky Survey (SDSS) \cite{york}, which launched in 2000 and provided recorded imagery of more than 35\% of the sky. Its current fourth phase that includes the study of the structure of nearby galaxies as well as our own is expected to run until 2020. A more recent example is the Gaia mission \cite{gaia} launched in 2013 and already released a map of the entire Milky Way galaxy \cite{gaia2}. These are two popular examples (amongst many) of collaborations that produce astronomical data sets of accelerated growth in both size and complexity. Future surveys are under development and many more are set to launch during the 2020's. An example is the Large Synoptic Survey Telescope (LSST) \cite{lsstb,lsst} with first light predicted in 2020 and data to be made public as soon as it is recorded.

If we look closely at the numbers expected to be produced by the LSST, we gain a better idea why ML algorithms are needed in Astronomy more than any time before. From exploring the transient sky to mapping the Milky Way, the LSST will photograph the entire available sky (almost all of the southern sphere) every 3 days for an entire decade commencing in 2022. It will produce raw data of about 15 TB per night, and about 60 PB in its lifetime. Furthermore, it will provide imagery of about 20 billion galaxies and a similar number of stars \cite{lsst}. The numbers for expected observations and recorded objects are enormous (billions or even trillions) and this will offer great opportunities for scientific discovery in many areas in Astronomy and observational Cosmology. However, conventional methods will not be enough to unleash the full potential of the LSST. This is not merely due to the expected size of the recorded data sets; rather the richness and complexity of the recorded imagery will require robust, sensitive and \textit{intelligent} tools for analyzing the data. New image processing methods, and specifically ML algorithms such as CNNs (which have an excellent reputation for such task in terms of performance vs. efficiency) need to be developed to accommodate the incoming enormous flow of recorded information. For example, as the number of recorded objects in an image grow larger, conventional classification methods (or even simple ML methods such as SVM) are more likely to fail. This prompts the need for a more sophisticated tool tailored towards complex data sets. The interdisciplinary development of ML algorithms well suited for astronomical imagery is key to harnessing the full extent of the potential offered by the LSST. This will transform how we interpret observations and will maximise scientific discovery as we enter the new era of Astronomy.

\section{Dynamical mass measurements of galaxy clusters}
Different ML algorithms are used for different purposes in Astronomy and in what follows, we summarise the findings of a recent study in 2019 by Ho et al. \cite{Ho}. The study used mock data to predict the dynamical mass of galaxy clusters, which is a galaxy-based method for estimating the abundance of galaxy clusters in the universe using the line-of-sight velocity of different galaxies in the cluster. Although not entirely intuitive, this information is vital with regard to constraining cosmological models as well as a powerful tool for dark matter detection. The population of a galaxy cluster is also important for determining the large-scale structure of the universe, and for galaxy evolution studies \cite{rev}. However, such measurement is highly sensitive to outliers (such as unbounded interloper galaxies) and discontinuities in the data collected (signaling an incomplete sample as we shall discuss in Section \ref{res}); and hence a robust tool for analyzing the observed data that produces as little scatter as possible is of high importance. The authors used two CNN models and their findings were shown to significantly reduce scatter in cluster mass estimation. The predictions of both CNN models were compared to traditional $M-\sigma$ methods, which infers the mass from the galaxy velocity dispersion $\sigma_v$, as well as a ML algorithm for dynamical mass estimation developed by Ntampaka et al. \cite{ntamp11,ntamp2}. The latter uses Support Distribution Machines (SDMs) which were shown to perform better than traditional methods in terms of reducing scatter (factor of 2).
\subsection{Conventional methods and motivation to go beyond}
Let us first discuss briefly the classical $M-\sigma$ method. This method relates the galaxy velocity dispersion $\sigma_v$ to the mass of the cluster in terms of a power law. However, this is an idealistic estimate and hinges on many assumptions often violated in reality. It assumes no interloper galaxies as well as spherical symmetry, which is not the case in real observations. It also assumes gravitational equilibrium which is easily disturbed by mergers. Nevertheless, the method has been successfully used to detect dark matter in the past (Coma cluster, 1933). Furthermore, it is used modernly along different methodologies to account for the presumed assumptions, such as interloper removal schemes. The power law in this scheme relating the velocity dispersion and the cluster mass in the context of this study is given by:
\begin{equation}
\label{virial}
\sigma_v=\sigma_{v,15}\left(\dfrac{h(z)M_{200c}}{10^{15}M_{\odot}}\right)^{\alpha}
\end{equation}
where $M_{200c}$ is the mass of the (assumed) spherical cluster and $h(z)$ is the dimensionless Hubble parameter. Furthermore, $\sigma_{v,15}$ and $\alpha$ are free parameters fixed by simulation and setting them to best-fit parameters. In \cite{Ho}, the $M-\sigma$ method was applied to two catalogs of mock data. The first is called the pure catalog which has no interloper galaxies and designed to mimic optimal interloper removal schemes. The second catalog is called the contaminated catalog which contains interloper galaxies, and the two findings are used to provide lower and upper limits on real $M-\sigma$ scatter.

The motivation for going beyond $M-\sigma$; replacing it by a more powerful and accurate ML algorithm is easy to find and we follow the footsteps of the discussion provided by the initial SDM paper \cite{ntamp11} in explaining why. $M-\sigma$ is dictated by a summary statistic $\sigma_v$, and this is manifest by the viral theorem's power law in equation \eqref{virial}. This simplification comes at the cost of neglecting important information in the line-of-sight velocity distributions of the galaxies within the cluster. Therefore when employing this method to make mass estimates, we are trading a degree of error and bias for simplicity and efficiency. However, what if we could make use of the information we are neglecting; utilize line-of-sight velocity distributions fully, cleverly overcome complications of triaxiality, environment, galaxy selection, and mergers, and do so efficiently? This is exactly where ML algorithms come into play. The mentioned factors inevitably produce high levels of scatter in the predictions made by $M-\sigma$; as the correlation between the velocity dispersion $\sigma_v$ and predicted mass is tainted. Alternatively, deducing the cluster mass from the line-of-site velocity distributions and making use of all information effectively is what ML algorithms are capable of achieving.
\subsection{ML methods}
\begin{figure}
\includegraphics[width=95mm]{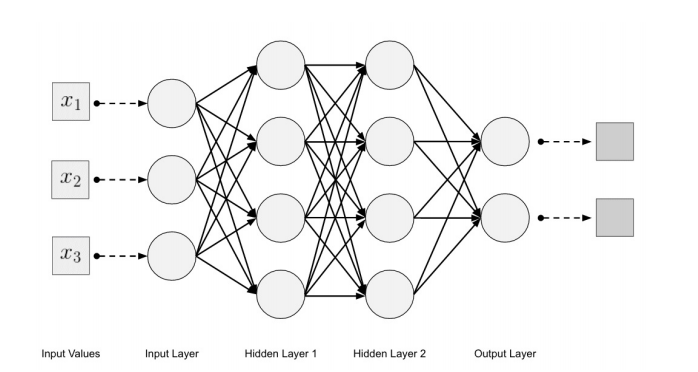}
\centering
\caption{\footnotesize An example of a feed-forward DNN with 2 hidden layers each consisting of 4 neurons. The input in this case can be described by 3-dimensional vector $\mathbf{x}^{(0)}$, while the output is a 2-dimensional vector $\mathbf{x}^{(3)}$. The connections between the neurons in different layers are characterized by weight matrices $\{\mathbf{W}^{(n)}\}$ and non-linear activation functions $\{f_n\}$. This figure was taken from \cite{book1}.}
\label{fig1}
\end{figure}

We now give a brief description of a CNN, which is a type of a feed-forward deep neural network (DNN) that uses convolution in at least one of its layers. DNNs are supervised ML algorithms used to predict a set of outputs through non-linearly complex relationships within a set of inputs. They consist of neurons that are related to each other by means of matrix multiplication and non-linear activation functions. That is, each neuron in some layer of the network hold a value that is a linear combination of the values of the neurons in the preceding layer, subject to an action of a non-linear activation function. We can describe the values at each layer in the network by:
\begin{equation}
\mathbf{x}^{(n)}=f_n\left(\mathbf{W}^{(n)}\cdot \mathbf{x}^{(n-1)}\right)
\end{equation}
where $n=1,2,...N-1$ such that $\mathbf{x}^{(0)}$ is the input vector and $\mathbf{x}^{(N)}$ is the output vector, $\{\mathbf{W}^{(n)}\}$ are weight matrices and $\{f_n\}$ are non-linear activation functions (e.g. sigmoid or ReLU). Figure \ref{fig1} shows an example where in this case, we have $\mathbf{x}^{(0)}$ being a 3-dimensional vector, $\mathbf{x}^{(3)}$ is a 2-dimensional vector, and the hidden layers are described by 4-dimensional vectors $\mathbf{x}^{(1)}$ and $\mathbf{x}^{(2)}$ respectively.
A DNN is characterized by its architecture which is roughly described by the number of neurons, number of layers and the connections between them. The training of a DNN aims to optimize the weight coefficients encoded in $\{\mathbf{W}^{(n)}\}$ using labeled data (with previously known output). This is usually achieved by means of back-propagation which uses Stochastic Gradient Descent to minimize a loss function that captures deviation of the prediction of the DNN from the correct output.

A CNN as mentioned previously has its name for the use of convolution in one or more of the layers. Mathematically, a convolution is a relationship between two functions that we denote by $\ast$. If $f$ and $g$ are real valued functions over $x$, then we define:
\begin{equation}
c(x)\equiv(f\ast g)(x)=\int f(t)g(x-t)dt,
\end{equation}
and in ML applications, we call the function $g$ the kernel and the function $f$ the input, and for practical purposes they are discretized. The output of the convolution is called the feature map, and generally, the input and kernel are multi-dimensional, that is:
\begin{equation}
\begin{aligned}
&c(i_1,i_2,i_3,...,i_N)\equiv(f\ast g)(i_1,i_2,i_3,...,i_N)
\\
&= \sum_{m_1}\sum_{m_2}\sum_{m_3}\cdots\sum_{m_N} f(m_1,m_2,m_3,...,m_N)g(i_1-m_1,i_2-m_2,i_3-m_3,...,i_N-m_N),
\end{aligned}
\end{equation}
and hence $f\ast g$ can be viewed as a weighted sum of the values of the function $f$ as dictated by the weight function $g$.

To put this into context, by performing such operation in convolution layers of a CNN, it can be roughly viewed as employing the usual weight matrix method in a generic DNN but with most entries set to zero. In other words, CNNs rely on minimal interactions between neurons which is achieved by making the kernel smaller than the input, this in turn makes it more efficient than matrix multiplication to perform the same task. In a generic feed-forward DNN, each output is connected to all preceding inputs, while in a CNN, each output is connected to a subset of preceding inputs called the receptive field. The combination of all receptive fields allow deeper layers to be indirectly connected to more elements in the input. Also, a CNN has tied weights in contrast to a generic DNN that uses weight matrices; this means the weight coefficients in a CNN (which are values of the kernel) can be used at all inputs in the layer which makes it more feasible in terms of memory usage and allows a feature to be detected in all receptive fields. A convolutional layer is usually followed by a pooling layer that reduces the dimensionality of the feature maps, and the operation used in \cite{Ho} is max pooling which simply passes only the maximum activation (obtained from the different receptive fields) to the next layer.

Two CNN models were used in \cite{Ho}, each has two convolutional layers but differ in the first layer to account for differences in the input used for each model. The reason for using CNNs at all for this particular task can be drawn from previous studies as well as CNNs reputation when trained as a regression. As mentioned earlier, SDM models were used for making dynamical mass predictions over mock catalogs \cite{ntamp2}, and more recently, simple regression models were used to mimic the performance of the SDM models \cite{ml1}. More complex ML algorithms were employed in \cite{ml2} but the error reduction was still comparable to previous methods. This prompts the use of complex yet efficient and tractable ML algorithms for the task at hand in order to achieve significant improvement. CNNs are known for these properties which is why they are very popular in Astronomy where data sets can be rich and complex. As we will discuss later, non-parametric algorithms are not best suited for interpreting complex data sets as their training sample size grows exponentially as the input is made more abundant. Alternatively, as we shall see in the next section, making the input data set more complex by adding additional information to the line-of-site velocity distributions (distance to cluster core in this case) does in fact improve the performance of CNN models significantly. This is due to the ability of CNN models to learn features given a relatively sparse input.

We will refrain from discussing the technicalities in the CNN models used for the purposes of this essay, and instead we turn to the results of using such algorithms in estimating the dynamical mass of galaxy clusters.
\subsection{Performance of CNN and comparative models}
\label{res}
\begin{figure}
\begin{floatrow}
\ffigbox{%
  \includegraphics[width=65mm]{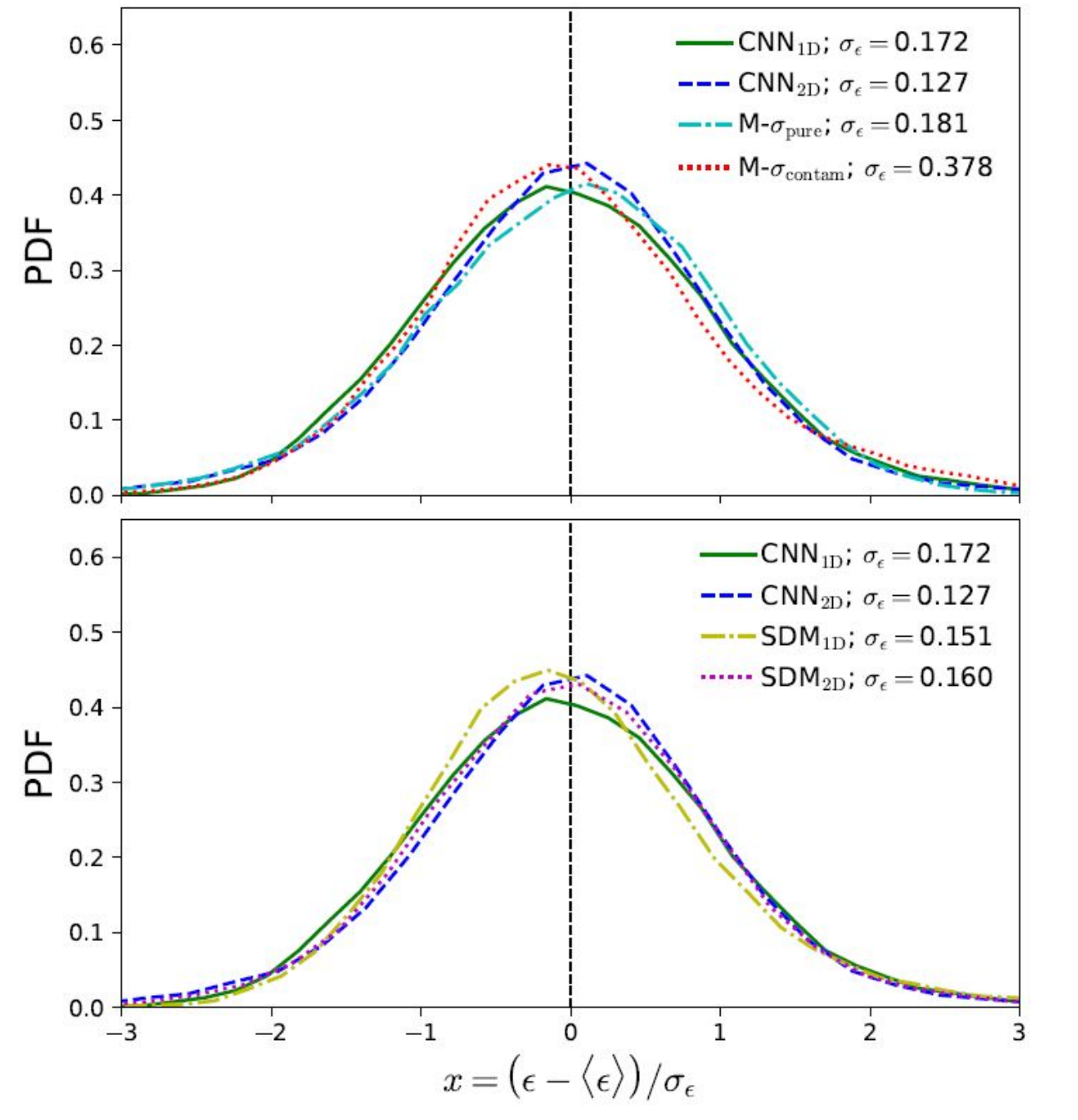}
}{%
  \caption{\footnotesize Distribution functions (PDFs) of normalised prediction residuals $x$ marginalized over true mass for all models considered. Non-Gaussianity in the PDFs are an indicator of systematic uncertainty produced by the model prediction. The upper panel shows comparison between CNN models of \cite{Ho} with conventional $M-\sigma$ models, while the lower panel shows comparison between the CNN models and SDM models of \cite{ntamp2}. The PDFs indicate that $\text{CNN}_{\text{2D}}$ produces the least systematic uncertainty. This figure is taken from \cite{Ho}.}%
  \label{fig2}
}
\capbtabbox{
  \tiny
  \begin{tabular}{ccccc}
  \hline
  Model & $\tilde{\epsilon}\pm \Delta \epsilon$ & $\sigma_\epsilon$ & $\gamma$ & $\kappa$ \\
  \hline
  \\
  \vspace{3mm}
  $\text{CNN}_{\text{1D}}$ & $0.011^{+0.169}_{-0.163}$&0.171&0.172&0.776 \\
  \vspace{3mm}
  $\text{CNN}_{\text{2D}}$ & $0.034^{+0.114}_{-0.119}$&0.127&0.097&1.577 \\
  \vspace{3mm}
  $M-\sigma_{\text{pure}}$ & $0.025^{+0.169}_{-0.183}$&0.181&-0.254&0.418 \\
  \vspace{3mm}
  $M-\sigma_{\text{contam}}$& $-0.087^{+0.365}_{-0.329}$&0.378&0.356&0.612 \\
  \vspace{3mm}
  $\text{SDM}_{\text{1D}}$ & $-0.060^{+0.146}_{-0.128}$&0.151&0.797&2.194 \\
  \vspace{3mm}
  $\text{SDM}_{\text{2D}}$ & $-0.014^{+0.148}_{-0.148}$&0.160&0.337&2.129 \\ \hline
  \end{tabular}
  \normalsize
}{%
  \vspace{15mm}
  \caption{\footnotesize A summary of performance characterization parameters for all models considered. $\text{CNN}_{\text{1D}}$ and $\text{CNN}_{\text{2D}}$ are the models developed in \cite{Ho} while $\text{SDM}_{\text{1D}}$ and $\text{SDM}_{\text{2D}}$ were developed in \cite{ntamp11,ntamp2}. Conventional $M-\sigma$ models are used to offer upper and lower bounds on scatter. All models used the contaminated catalog (includes interloper galaxies) except $M-\sigma_{\text{pure}}$ which used an idealized catalog based on interloper-removal methods to offer a lower bound on scatter. Values in this table were taken from \cite{Ho}.}
  \label{tab1}%
}
\end{floatrow}
\end{figure}

In this section, we discuss and analyze the results presented in \cite{Ho} of employing CNN models to dynamical mass measurements of galaxy clusters. To characterize the performance of each CNN model used, the logarithmic residual $\epsilon$ was calculated which is defined as:
\begin{equation}
\epsilon=\mathrm{log}_{10}\left(\dfrac{M_{\text{predicted}}}{M_{\text{true}}}\right),
\end{equation}
as well as the median $\tilde{\epsilon}$ over cumulative statistics of the $\epsilon$ distribution, the 16-84 percentile range $\Delta\epsilon$, and standard deviation scatter $\sigma_\epsilon$. These values for different models employed are shown in Table \ref{tab1}. The subscripts 1D and 2D indicate different input data sets, the former only contains line-of-site velocities and is one dimensional, while the latter contains further information with regard to the radial distance to the cluster centre, and is 2-dimensional. Note that Table \ref{tab1} also contains values for the skewness $\gamma$ and kurtosis $\kappa$. $\gamma$ and $\kappa$ are relevant in this analysis for estimating the bias based on cluster counts.
This is done by employing an Edgeworth expansion at fixed red-shift for the observable mass relation, that is, if $x=(\epsilon-\langle\epsilon\rangle)/\sigma_\epsilon$ is the normalized logarithmic residual ($\langle \epsilon \rangle$ is the mean) and $G$ is the Gaussian distribution, then we have the distribution function relating $M_{\text{predicted}}$ and $M_{\text{true}}$ \cite{Ho}:
\begin{equation}
P\left(M_{\text{predicted}}|M_{\text{true}}\right)\approx G(x)-\dfrac{\gamma}{6}\dfrac{d^3G}{dx^3}+\dfrac{\kappa}{24}\dfrac{d^4G}{dx^4}
\end{equation}
where $\gamma$ and $\kappa$ are the skewness and kurtosis of the $x$ distribution, respectively. Since the cluster abundance follows a power law (in mass) with power $\alpha$ (2 in this case), then the predicted cluster abundance can also be written as \cite{Ho}:
\begin{equation}
\dfrac{d n}{d \mathrm{ln}M_{\text{predicted}}}\approx \left(\dfrac{d n}{d \mathrm{ln}M_{\text{predicted}}}\right)_0 \left(1+\dfrac{\alpha^3\sigma^3\gamma}{2}+\dfrac{\alpha^4\sigma^4\kappa}{24}\right)
\end{equation}
where $\left(d n/d \mathrm{ln}M_{\text{predicted}}\right)_0$ is the cluster abundance of a purely log-normal $x$ distribution. This equation is used to estimate the systematic uncertainty produced by each model. In short, what this means is that non-Gaussianity in the predictions made by a model can produce bias in the estimation. Hence, the more Gaussian a PDF (in $x$) for a given model the less impact non-Gaussian uncertainty has on abundance measurements. We see clearly from Figure \ref{fig2} that $\text{CNN}_{\text{2D}}$ predictions produce the least systematic uncertainty (1.1\%), and both $\text{CNN}_{\text{1D}}$ (4.2\%) and $\text{CNN}_{\text{2D}}$ produce less systematic uncertainty than $\text{SDM}_{\text{1D}}$ (6.5\%) and $\text{SDM}_{\text{2D}}$ (4.9\%), and these are compared to the value produced by $M-\sigma_{\text{pure}}$ which is 1.6\%. Hence we see that $\text{CNN}_{\text{2D}}$ does in fact approach an idealised measurement and exceeds the performance of conventional interloper-removal procedures (note that for $M-\sigma_{\text{contam}}$ the systematic uncertainty is at 55\% due to its high scatter). The exact numerical values were taken from \cite{Ho}.

All models were evaluated using the contaminated catalog mentioned earlier, except $M-\sigma_{\text{pure}}$ which used the pure catalog to serve as a lower bound. Since the input data set for the model $\text{CNN}_{\text{2D}}$ has more information than that of $\text{CNN}_{\text{1D}}$, the predictions of $\text{CNN}_{\text{2D}}$ exhibit significantly less scatter than those of $\text{CNN}_{\text{1D}}$ (0.127 dex compared to 0.171 dex for $\sigma_\epsilon$ implies 25\% reduction in scatter \cite{Ho}). Interestingly, this intuitive result does not hold for SDM models \cite{ntamp2}; as the additional information in the input data set for $\text{SDM}_{\text{2D}}$ does not lead to lower scatter. We see in fact in Table \ref{tab1} that the predictions of $\text{SDM}_{\text{2D}}$ exhibit more scatter than those of $\text{SDM}_{\text{1D}}$. This is because SDMs non-parametric structure does not allow them to probe complex training data sets in order to compare new clusters; since as training data sets become more complex, the requirements for SDM training sample size grow exponentially. This is not the case for CNNs since as mentioned earlier, the use of convolution layers will limit the input parameters (by means of a receptive field) and a CNN is capable of learning over more complex data sets without any obstruction. Albeit for poorly understood reasons, this property does make it advantageous to favor the use of CNNs in future surveys.

Another point worth mentioning is that the ML models $\text{SDM}_{\text{1D}}$ and $\text{SDM}_{\text{2D}}$ first developed in \cite{ntamp11,ntamp2} do reduce scatter below that of the idealized $M-\sigma_{\text{pure}}$ and even $\text{CNN}_{\text{1D}}$, however, they also produce high bias in the estimation. As reported in \cite{Ho}, $\text{SDM}_{\text{1D}}$ under-predicts low mass clusters while $\text{SDM}_{\text{2D}}$ under-predicts high mass clusters. This can be problematic when one is interested in precision measurements where the model used should be trustworthy. This is not the case when compared to $\text{CNN}_{\text{2D}}$ which produces both lower scatter and bias. When comparing CNN models used to conventional $M-\sigma$ models, $\text{CNN}_{\text{1D}}$ and $\text{CNN}_{\text{2D}}$ reduce scatter by 55\% and 66\% when compared to $M-\sigma_{\text{contam}}$, and by 6\% and 30\% when compared to $M-\sigma_{\text{pure}}$, respectively. The better SDM model used is $\text{SDM}_{\text{1D}}$ and when compared to $\text{CNN}_{\text{2D}}$, the performance of $\text{CNN}_{\text{2D}}$ shows a reduction in scatter by 20\%. This suggests that CNN models and specifically $\text{CNN}_{\text{2D}}$ can predict the dynamical mass of galaxy clusters both better than conventional methods, and other modern ML algorithms presented in recent literature.

A further important characterization of the performance of different models considered for dynamical mass estimation is their robustness under variations in the sampling rate. What this means is that we do not only seek an accurate model to estimate the mass of perfectly constructed fully sampled galaxy clusters, but we also want our model to work just as accurately when galaxies are randomly removed from the cluster, that is if the input is a sub-sample of the galaxies in the cluster. This is important since in reality, different galaxies in the cluster can be indistinguishable or unobservable and it translates into the sample being incomplete. To account for such inevitability, sub-sampled mass deviation $\epsilon^{(r)}$ was used in order to capture the deviation for each model from the correct prediction when random galaxies are removed from the cluster. This is defined as:
\begin{equation}
\epsilon^{(r)}=\mathrm{log}_{10}\left(\dfrac{\bar{M}^{(r)}_{\text{predicted}}}{M^{(1.0)}_{\text{predicted}}}\right)
\end{equation}
where $r \in [0,1]$ denotes the rate of galaxies randomly sub-sampled without replacement, $\bar{M}^{(r)}_{\text{predicted}}$ is the average sub-sampled mass prediction calculated from 10 sub-sampled combinations, and $M^{(1.0)}_{\text{predicted}}$ is the fully sampled prediction. The reason the selection was random is to avoid newly introduced dependencies in the measurements, and since there are many combinations that can be used to choose galaxies from a cluster, an average was taken over 10 combinations for each cluster \cite{Ho}. If a model exhibits strong dependency on cluster richness in its mass prediction, then this will manifest itself via a strong correlation between $r$ and $\epsilon^{(r)}$. Note that this does not mean the model is not accurate; the prediction made by the model for a given cluster might be more accurate than other models considered, however, it signals possible strong deviation from the correct prediction given an unconstrained sampling rate. This is a measure of how trustworthy a model is if not all galaxies in the cluster were accounted for in the input data set, i.e. a measure of robustness given a degree of ignorance.

A cumulative statistic was constructed for $r \in [0.6,0.8]$ that describes the median $\tilde{\epsilon}^{(0.6-0.8)}$ and 16-84 percentile scatter $\Delta \epsilon^{(0.6-0.8)}$. That is, the sampling rate is allowed to change from 60-80\%, which is a realistic range, and $\tilde{\epsilon}^{(0.6-0.8)}$ captures the bias (richness dependence) while $\Delta \epsilon^{(0.6-0.8)}$ captures the scatter in such range, within the models' prediction. Hence, ideally one seeks a model that minimizes both $|{\tilde{\epsilon}^{(0.6-0.8)}}|$ and $\Delta \epsilon^{(0.6-0.8)}$, which roughly indicates that such model has minimal dependency on sampling rate (it can produce accurate predictions given any sub-sample of the cluster), and we are confident that it does not deviate strongly from this minimal dependency on $r$ behavior given an arbitrary number of measurements, which is desirable for realistic applications. Values for $\tilde{\epsilon}^{(0.6-0.8)}$ and $\Delta \epsilon^{(0.6-0.8)}$ for the different models considered are shown in Table \ref{tab2}.

\begin{wrapfigure}{R}{0.58\textwidth}
\begin{floatrow}
\capbtabbox{
  \tiny
  \begin{tabular}{ccc}
  \hline
  Model & $\tilde{\epsilon}^{(0.6-0.8)}$&$\Delta \epsilon^{(0.6-0.8)}$ \\
  \hline
  \\
  \vspace{3mm}
  $\text{CNN}_{\text{1D}}$ &-0.029&0.113 \\
  \vspace{3mm}
  $\text{CNN}_{\text{2D}}$ &-0.028 &0.082 \\
  \vspace{3mm}
  $M-\sigma_{\text{pure}}$ &0.004  &0.237\\
  \vspace{3mm}
  $M-\sigma_{\text{contam}}$& 0.001 &0.269\\
  \vspace{3mm}
  $\text{SDM}_{\text{1D}}$ &-0.140  &0.174\\
  \vspace{3mm}
  $\text{SDM}_{\text{2D}}$ &0.001  &0.255\\ \hline
  \end{tabular}
  \normalsize
}{%
  \caption{\footnotesize Cumulative statistics of robustness under sampling rate variation for the different models considered in Table \ref{tab1}. Values were taken from \cite{Ho}.}
  \label{tab2}%
}
\end{floatrow}
\end{wrapfigure}
We see from Table \ref{tab2} that the CNN models are the most trustworthy since they exhibit lowest values in scatter for $r\in [0.6,0.8]$. This is a 53\% improvement in reducing the residual ranges for $\text{CNN}_{\text{2D}}$ when compared the best of $M-\sigma$ and SDM models, which is $\text{SDM}_{\text{1D}}$ \cite{Ho}. This robust behavior of CNN models under variations in $r$ is an important property to consider when making realistic predictions, and might indicate that CNNs are more suitable for precision measurements.

Although less crucial than prediction error comparisons, CNN models are significantly more computationally efficient than SDM models. This can be relevant when high-quality data sets are abundant and in realistic applications where input data sets are much larger than the mock catalogs used. For CNN models discussed in this essay, the full time taken in training and evaluation was 10 minutes, compared to 6 hours for SDM models \cite{Ho}. This points to CNN models being more suitable for realistic applications in terms of efficiency and tractability.

The above discussion of the results presented in \cite{Ho} does suggest that CNNs can significantly offer more accurate and robust mass predictions for galaxy clusters than any other method considered in the literature. However, this would be a quick and naive conclusion to make at this stage since there are factors that could come into play when considering other input data sets than the ones used. We elaborate on this point in Section \ref{diss}.
\section{Other triumphs and future prospects}
In the previous section, we discussed the work done by Ho et al. \cite{Ho} and we saw clearly that CNN models outperform conventional and other modern ML algorithms with regard to estimating the dynamical mass of galaxy clusters. However, this is merely the tip of the iceberg; as for a variety of tasks in Astronomy, ML algorithms were shown to outperform conventional statistical models. In this section, we mention two more studies where DNN models were successfully tested and shown to exceed the limitations of conventional means. We also give a general outline of what the future of astronomical data looks like, and in this light, how the fields of ML and Astronomy can benefit from upcoming large scale surveys.

As mentioned in the introduction, CNNs are widely used in studying and detecting gravitational lensing; this is not a surprise as CNNs are particularly suitable for image recognition and classification tasks. It was shown in a study performed by Hezaveh et al. \cite{Hez} that CNN models can estimate lensing parameters up to 10 million times faster than conventional Maximum Likelihood methods. This makes them a more efficient replacement to conventional procedures, especially as astronomical data sets become more rich and complex.

More recently, a DNN model was used to perform an $N$-body simulation of the universe \cite{He}. This is done to predict the large scale structure of our universe using comparative data collected from different sky surveys. Conventional $N$-body simulation methods are not computationally efficient as they tend to be very expensive. The DNN model used in \cite{He} was shown to outperform traditional second order perturbation theory and has the ability to extrapolate beyond its training date, i.e. produce accurate predictions given different cosmologies.

It is a fact that Astronomy is entering an era of big data, since large sky surveys such as the LSST and radio interferometers such as the Hydrogen Epoch of Re-ionization Array (HERA) are set to produce data at \textit{truly astronomical} rates. Many other projects are due to launch during the next decade and as mentioned earlier, traditional methodologies are simply inadequate in effectively analyzing the expected data sets. In this respect, the following decade will be a great opportunity for Astronomers and ML experts alike to work hand in hand on this revolutionary interdisciplinary development of robust techniques tailored towards big astronomical data. This will offer a real life testing of various ML algorithms that will accelerate our understanding and development in such field, as well as maximise the utility of our large, complex and rich data sets in helping us understand our universe.

\section{Summary and discussion}
\label{diss}
As astronomical data sets grew both in size and complexity, ML algorithms proved to excel at different tasks in analyzing, classifying and characterizing the data collected from various surveys. In some cases ML algorithms were tailored towards an application in Astronomy (see Probabilistic Random Forest for example \cite{PRF}) and outperformed conventional methods in terms of accuracy, efficiency and predictability. In this essay we reviewed successes of employing deep learning algorithms in dynamical mass estimation of galaxy clusters, in particular, we focused on the work done by Ho et al. \cite{Ho}. In Section \ref{res}, we discussed the performance of two CNN models used to predict the dynamical mass of galaxy clusters using mock catalogs. The CNN models outperform conventional $M-\sigma$ methods as well as modern SDM models \cite{ntamp2} in terms of reducing scatter, accuracy, robustness and efficiency.

Although the CNN models used in \cite{Ho} passed a specific test with flying colors, the results are not generalisable for a variety of reasons. The use of mock catalogs in general produces caveats in the discussion proposed thereafter since the data used in such sets are more often than not unrepresentative of real life observations. This is the case in \cite{Ho} and we discuss the reasons for this in what follows.

First of all, the data used in \cite{Ho} is from the MultiDark Planck 2 simulation \cite{multd} only which fixes abundancy rates in clusters. This can be problematic when drawing general conclusion since SDM models clearly suffered from sparsity in the 2-dimensional input data set. It would have been more conclusive if the CNN models were shown to outperform the SDM models given an abundant sample. Although it would not be computationally efficient to use SDM in that case.

Another caveat of only using one source of data is the assumptions made for creating realistic mock catalogs. There are a number of assumptions made in \cite{Ho} for creating the contaminated catalog and the results for the performance of the CNN models might rely on these assumptions. These assumptions are summarised as setting specific mass bounds on what is a galaxy and what is a cluster, assuming prior knowledge of the cluster centre, and assuming a specific value of red-shift for the cluster ($z=0.117$) and observer ($z=0$). How exactly this affects the performance of the CNN models is absent in \cite{Ho} since there is no mention of more than one criteria used to create realistic mock observations, and hence no way to compare. This can be made more concrete by testing the models on unseen data sets created under different assumptions than the ones used in the training of the model, e.g. as done in \cite{ml2} to minimize the unrealistic attributes' effects of mock catalogs on the prediction.

When it comes to the contaminated catalogs of \cite{Ho} themselves, there is also room for improvement with regard to taking into account realistic effects in clusters in a direct manner. Effects such as cluster mergers, triaxiality, and interloper galaxies were accounted for. However, other important features that correlate with mass such as luminosity and group dynamics were not mentioned.

In general, the work done \cite{Ho} offers a first step towards making more accurate real life predictions for the dynamical mass of galaxy clusters. The need for generalisation and further strengthening the results is inevitable before drawing absolute conclusions. A real life input data set might contain just enough objects for SDM to outperform CNN models. To be able to address this, more generalizable tests need to be employed using a diverse family of data sets. Nevertheless, the future of using CNNs for real dynamical mass measurements, and precision cosmology in general, does seem very promising.


\end{document}